\documentclass[letter]{aa}

\usepackage{graphicx}
\usepackage{txfonts}
\usepackage[colorlinks=true,
            urlcolor=blue,
            linkcolor=blue,
            citecolor=blue]
            {hyperref}
\usepackage{tabu}
\usepackage{ulem}
\usepackage{multirow} 
\usepackage{gensymb}

\newcommand{\Ha}{H$\alpha$}
\newcommand{\Hb}{H$\beta$}

\newcommand{\mcc}[1]{\multicolumn{1}{c}{#1}}

\newcommand{\kms}{km\,s$^{-1}$}

\begin{document} 

   \title{
   Double-peaked \ion{Ca}{ii} traces a relativistic broad-line region disk in NGC\,4593
   }

      \author{M.W. Ochmann \inst{1},
           P.M.~Weilbacher\inst{2},
           M.A.~Probst \inst{1},
           W.~Kollatschny \inst{1},
           D.~Chelouche \inst{3,4},
           R.~Chini \inst{5,6,7},
           D.~Grupe \inst{8},
           M.~Haas \inst{5},
           S.~Kaspi \inst{9},
           S.~Komossa \inst{10}      
           }

   \institute{
          Institut f\"ur Astrophysik und Geophysik, Universit\"at G\"ottingen,
          Friedrich-Hund Platz 1, 37077 G\"ottingen, Germany \\
          \email{martin.ochmann@uni-goettingen.de}
          \and
          Leibniz-Institut f\"ur Astrophysik Potsdam (AIP), An der Sternwarte 16, 14482 Potsdam, Germany
          \and
          Department of Physics, Faculty of Natural Sciences, University of Haifa, Haifa 3498838, Israel
          \and
          Haifa Research Center for Theoretical Physics and Astrophysics, University of Haifa, Haifa 3498838, Israel
          \and
          Ruhr University Bochum, Faculty of Physics and Astronomy, Astronomical Institute (AIRUB), 44780 Bochum, Germany
          \and
          Nicolaus Copernicus Astronomical Center, Polish Academy of Sciences, Bartycka 18, 00-716 Warszawa, Poland
          \and
          Universidad Católica del Norte, Instituto de Astronomía, Avenida Angamos 0610, Antofagasta, Chile
          \and
          Department of Physics, Geology, and Engineering Technology, Northern Kentucky University, 1 Nunn Drive, Highland Heights, KY 41099, USA
          \and
          School of Physics \& Astronomy and the Wise Observatory, The Raymond and Beverly Sackler Faculty of Exact Sciences, Tel-Aviv University, Tel-Aviv 6997801, Israel
          \and
          Max-Planck-Institut für Radioastronomie, Auf dem Hügel 69, D-53121 Bonn, Germany
          }

  \date{Received 27 March 2025 / Accepted 16 April 2025}

 \abstract{Double-peaked emission lines are observed in a small percentage of active galactic nuclei (AGN). These lines allow the determination of fundamental properties of the line-emitting region, known as the broad-line region (BLR).
}
{We investigated the structure and kinematics of the BLR in the nearby Seyfert galaxy NGC\,4593 through an analysis of the near-infrared (NIR) line blend of \ion{Ca}{ii}$\,\lambda8498,\lambda8542,\lambda8662$, and \ion{O}{i}$\,\lambda8446$ observed in a 2019 VLT/MUSE spectrum.
}
{We performed a detailed decomposition of the NIR \ion{Ca}{ii} triplet and \ion{O}{i}$\,\lambda8446$ blend, extracting clean profiles of  \ion{Ca}{ii}$\,\lambda8498,\lambda8542,\lambda8662$, and \ion{O}{i}$\,\lambda8446$. We then fitted \ion{Ca}{ii}$\,\lambda8662$ with a relativistic elliptical line-emitting accretion disk model.
}
{The extracted line profiles are double-peaked with a full width at half maximum (FWHM) of $\sim 3700$\,\kms{} and exhibit a redward asymmetry with a red-to-blue peak ratio of 4:3. The \ion{Ca}{ii} triplet lines have an intensity ratio of 1:1:1 and show no evidence of a central narrow or intermediate-width component. The profiles of \ion{Ca}{ii} and \ion{O}{i} are remarkably similar, suggesting a common region of origin. Given the 1:1:1 ratio of the \ion{Ca}{ii} triplet, this region is likely a  high-density emission zone, and the \ion{Ca}{ii}$\,\lambda8662$ profile is well described by a mildly eccentric, low-inclination relativistic line-emitting disk with minimal internal turbulence. The profile represents one of the clearest kinematic signatures of a relativistic disk observed in BLR emission lines to date.
}
{The double-peaked profiles of the NIR \ion{Ca}{ii} triplet and \ion{O}{i}$\,\lambda8446$ in NGC\,4593 represent the first detection of double-peaked \ion{Ca}{ii} and \ion{O}{i}$\,\lambda 8446$ in a nontransient AGN spectrum. The minimal intrinsic turbulence (the lowest value reported for an AGN emission line to date) and the absence of narrow or intermediate-width components in \ion{Ca}{ii}$\,\lambda8662$ make it a powerful diagnostic tool of BLR structure and kinematics. Further investigations of the profiles of \ion{Ca}{ii} and \ion{O}{i} in other AGN are recommended to better constrain BLR properties and the nature of the underlying accretion flow.
}

  \keywords{galaxies: active - galaxies: Seyfert – galaxies: nuclei – quasars: individual: NGC 4593 – quasars: emission lines}

  \titlerunning{Double-peaked \ion{Ca}{ii} in NGC\,4593}
  \authorrunning{M.W. Ochmann et al.}

  \maketitle
%
%

%
\section{Introduction}\label{sec:introduction}
%
Broad emission lines in the spectra of active galactic nuclei (AGN) are  powerful tools for probing the structure and kinematics of the gas in the immediate vicinity (within a few thousand gravitational radii, $r_{\rm g}$) of the central supermassive black hole (SMBH). However, the profiles of broad emission lines in AGN are typically complex and exhibit a wide range of shapes and widths that vary from object to object  \citep[e.g.,][]{sulentic00}. These complex profiles may indicate rather complex broad-line region (BLR) kinematics (e.g., outflows or inflows); additional effects, such as contributions from narrow or intermediate components in the narrow-line region (NLR) or intermediate-line region \citep[IRL; e.g.,][and references therein]{popovic04, adhikari18}; as well as turbulence \citep[][]{kollatschny13a}, disk winds \citep[][]{murray97}, and optical depth effects \citep[][]{eracleous09, flohic12}, or a combination of all these factors. As a result, it is generally not possible to unambiguously disentangle and constrain the kinematic components shaping the profiles based on single-epoch spectra alone; this means that  for most AGN the underlying kinematic scenario cannot be directly inferred from the observed line profiles, and instead other methods such as reverberation mapping \citep[RM;][]{blandford82, kollatschny03, horne21}, interferometry \citep{gravity18}, or microlensing \citep[e.g.,][]{schmidt10,fian21}  have to be employed. However, RM campaigns are time-consuming and challenging to organize, and require extensive observational resources with homogeneous instrumentation \citep[e.g.,][]{sobrino25}, and interferometry and microlensing studies only work for a comparably small number of objects. Therefore, in the majority of AGN, the connection between line profiles and BLR kinematics remains ambiguous.

A notable exception to this general constraint are double-peaked emitters (DPEs), which are assumed to arise due to the kinematics of the BLR being dominated by the kinematics of a rotating relativistic line-emitting (accretion) disk \citep[e.g.,][and references therein]{chen89a,chen89b}. Only a small percentage of the general AGN population exhibit clear double-peaked profiles \citep[][]{strateva03, eracleous03, ward24} and such profiles occur predominantly in low-luminosity systems \citep[][]{elitzur14}. Several line-emitting disk models have been proposed, including circular \citep[][]{chen89b}, eccentric \citep[][]{eracleous95, storchi-bergmann97}, warped \citep[][]{wu10}, and spiral-structured disks \citep[][]{gilbert99, storchi-bergmann03} or disks with rotating hot spots \citep[e.g.,][]{newman97}. In particular, circular, elliptical, and spiral-arm models have been successfully used to constrain BLR structure and kinematics in DPEs \citep[e.g.,][]{eracleous03, strateva03, storchi-bergmann17, ward24}.

Previous DPE analyses have focused almost exclusively on Balmer lines (\Hb{} and \Ha{}), while studies of double-peaked UV \citep[e.g.,][]{bianchi22} or near-infrared (NIR) lines \citep[e.g.,][]{diasdossantos23, ochmann24} remain rare. More generally, DPE analysis is often complicated by narrow or intermediate-width components \citep[e.g.,][]{storchi-bergmann17}, and most observed profiles require significant intrinsic turbulence ($\gtrsim 500$\,\kms{}) to achieve a proper fit, effectively smearing out small-scale kinematic features. 

Here we report the detection of double-peaked NIR \ion{Ca}{ii} triplet and \ion{O}{i}$\,\lambda 8446$ lines in NGC\,4593  ($\alpha_{2000}=12^\text{h}39^\text{m}39.44^\text{s}, \delta_{2000} = -05^{\circ}20^{\prime}39.034^{\prime \prime}$), a local ($z=0.008312$) face-on Seyfert galaxy,\footnote{\url{https://ned.ipac.caltech.edu/}} extensively studied in past variability campaigns \citep[][]{dietrich94, santos-lleo95,kollatschny97,denney06,barth15,mchardy18, cackett18,chen22,kumari23}.  Spectra covering the \ion{Ca}{ii} triplet and \ion{O}{i} region in NGC\,4593 have previously been presented by \citet{garcia-rissmann05} and \citet{cackett18}.

%
\section{Observations}\label{sec:observations}
%

NGC\,4593  was observed with the VLT/MUSE \citep[Multi Unit Spectroscopic Explorer;][]{bacon10, bacon14} IFU spectrograph as part of  ESO Program 0103.B-0908 (PI: Knud Jahnke) on 2019 April 28. The observation was carried out in adaptive optics (AO) corrected narrow-field mode (NFM-AO-N) with an exposure time of 4800\,s in external seeing of $\sim$0\farcs8. The AO correction achieved a central peak $\lesssim$0\farcs09 in the $I$-band. MUSE covers the optical and NIR wavelength range between $\sim 4700$\,\AA\ and $9300$\,\AA\ at a spectral resolution of $\sim 2.5$\,\AA. The spectra are sampled at 1.25\,\AA\ in dispersion direction and at 0\farcs025 in spatial direction. The signal-to-noise ratio (S/N) in an aperture of 0\farcs20 radius exceeds 120 for the \ion{Ca}{ii} triplet. Details on the data reduction are given in Appendix~\ref{sec:app_data_reduction}.

%
\section{Results}\label{sec:results}
%

%
\subsection{The optical to near-infrared MUSE spectrum of NGC\,4593 }\label{sec:results_spectrum}
%

We show the optical to NIR MUSE spectrum extracted within a 0\farcs20-radius aperture, covering a rest-frame range of $4711$–$9272$\,\AA{}, in Fig.~\ref{fig:NGC4593_totalspectrum}. Details on the spectral extraction are given in Appendix~\ref{sec:app_spectral_extraction}. The spectrum is reddened and features numerous emission lines, including strong coronal line emission,\footnote{We observe complex coronal line profiles that appear split. A similar line splitting in coronal lines was reported by \citet[][]{mazzalay10} in several galaxies.} with all prominent lines identified. Broad emission appears in the Balmer lines, \ion{He}{i}, \ion{He}{ii}, and notably in the NIR calcium triplet + \ion{O}{i}$\,\lambda8446$ blend. A sharp double-peaked profile is seen in \ion{Ca}{ii}$\,\lambda8662$, with its blue wing blending into the red wing of \ion{Ca}{ii}$\,\lambda8542$; this feature is also discernible, though less distinctly, in the spectrum of NGC\,4593 presented by \citet{garcia-rissmann05}. No clear double-peaked profile is discernible in the Balmer or helium lines, though a prominent red shoulder in \Hb{} may indicate a double-peaked \Hb{} profile otherwise obscured by a narrow component.

\begin{figure*}[h!]
    \centering
    \includegraphics[width=1\textwidth,angle=0]{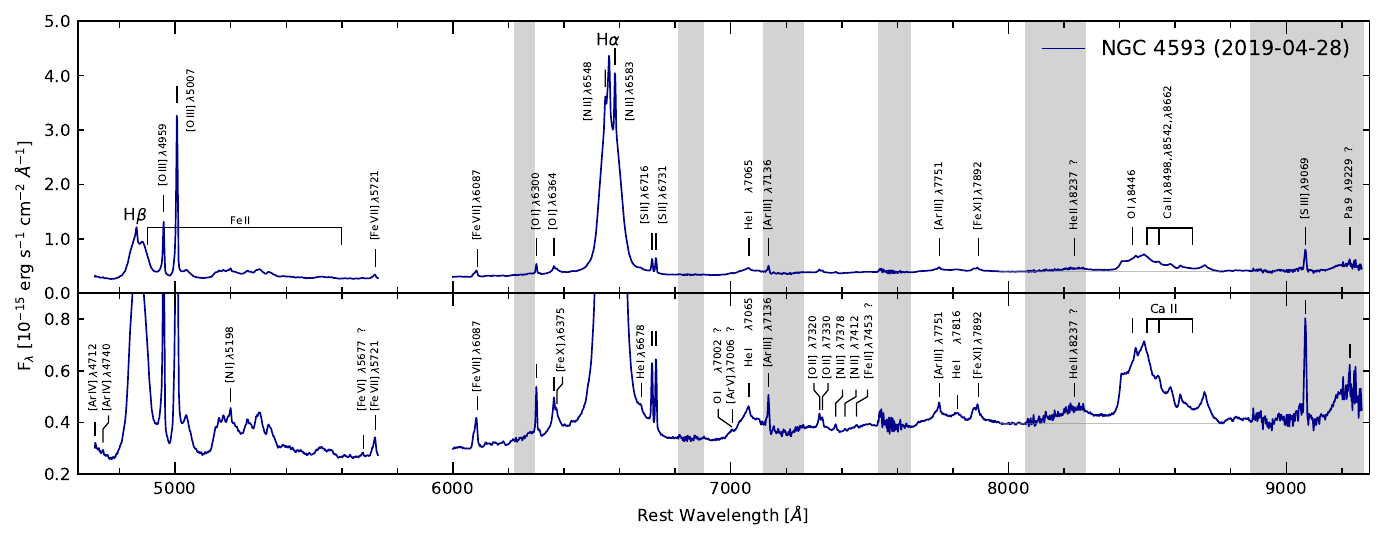}
    \caption{MUSE spectrum of NGC\,4593  extracted within a 0\farcs20-radius aperture obtained in NFM-AO-N on 28 April 2019. The upper panel shows the full spectrum with the most prominent narrow and broad emission lines labeled. To enhance weaker emission features, the lower panel shows a close-up of the spectral continuum. The linear pseudo-continuum used for further analysis of the \ion{Ca}{ii} triplet + \ion{O}{i}$\,\lambda8446$ complex is indicated by a gray line. Regions affected by telluric absorption bands are indicated by gray shading.}
    \label{fig:NGC4593_totalspectrum}
\end{figure*}

%
\subsection{Double-peaked \ion{Ca}{ii} triplet and \ion{O}{i}$\,\lambda 8446$ emission lines}\label{sec:results_extracting}
\begin{figure}[h!]
    \centering
    \includegraphics[width=0.46\textwidth,angle=0]{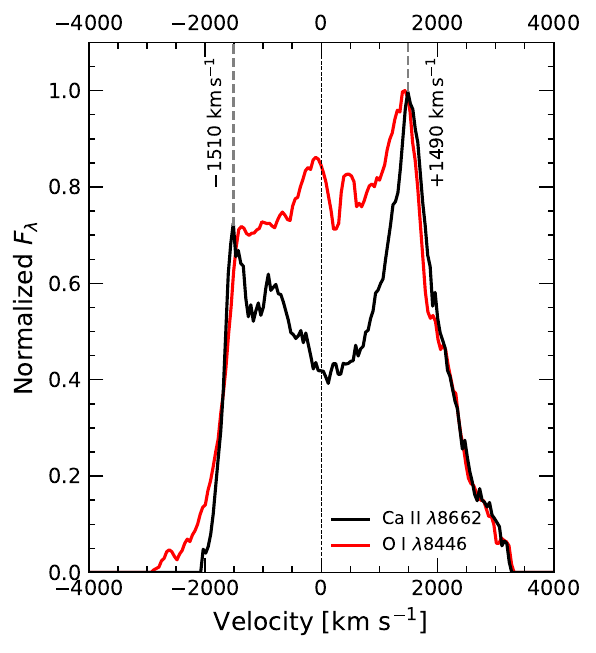}
    \caption{Comparison of the \ion{Ca}{ii}$\,\lambda8662$ (black) and \ion{O}{i}$\,\lambda8446$ (red) line profiles in velocity space. The central wavelength and peak velocities are marked by the black and gray dashed lines, respectively.}
    \label{fig:OI_CaII_comparison_NGC4593}
\end{figure}

%
We extracted clean profiles of \ion{Ca}{ii}$\,\lambda8662$ and \ion{O}{i}$\,\lambda8446$ through a detailed decomposition of the NIR \ion{Ca}{ii} and \ion{O}{i} blend (see Appendix~\ref{sec:app_decomposing_OI_CaII}). A direct comparison of the \ion{Ca}{ii}$\,\lambda8662$ and \ion{O}{i}$\,\lambda8446$ profiles is shown in Fig.~\ref{fig:OI_CaII_comparison_NGC4593}. Both profiles are double-peaked and exhibit a redward asymmetry with a red-to-blue peak ratio of 4:3. The peaks of the \ion{Ca}{ii}$\,\lambda8662$ profile are located at $-1510$\,\kms{} and $+1490$\,\kms{}, while the peaks of the \ion{O}{i}$\,\lambda8446$ profile are shifted toward the central wavelength by $\sim 60$-$130$\,\kms{}. The red wings of the \ion{Ca}{ii} and \ion{O}{i} profile are identical in shape with a distinct red ``foot'' present in both profiles. In contrast, the blue wing of \ion{O}{i}$\,\lambda8446$ is shallower than that of \ion{Ca}{ii}$\,\lambda8662$, similar to what was observed for NGC\,1566 \citep{ochmann24}.

The \ion{Ca}{ii} triplet lines exhibit identical profiles with a ratio of 1:1:1. Unlike \ion{O}{i}$\,\lambda8446$, which exhibits an additional central narrow and/or intermediate-width component that effectively masks the double-peaked nature of the profile, the \ion{Ca}{ii} triplet profiles show no evidence of such a feature. Instead, the \ion{Ca}{ii}$\,\lambda8662$ profile exhibits small peaks in the blue part of the profile (see \ref{sec:results_fitting} for details).

When  normalized to peak intensity at $v = 1500$\,\kms{}, the \ion{Ca}{ii} and \ion{O}{i} profiles have nearly identical line widths (see Table~\ref{tab:lineWidths}) with a full width at half maximum (FWHM$_{v=1500}$) of $3680$\,\kms{} and $3580$\,\kms{}, respectively. When normalized to the flux at $v = 0$\,\kms{}, the profiles have differing widths (FWHM$_{v=0}$) of $4360$\,\kms{} and $3830$\,\kms{} for \ion{Ca}{ii} and \ion{O}{i}, respectively. This can be attributed to the presence of the additional central narrow to intermediate-width component in \ion{O}{i}$\,\lambda8446$.

%

%
%
\begin{table}[!htb] 
\centering 
\caption{FWZI and FWHM of the \ion{Ca}{ii} and \ion{O}{i} profiles in NGC\,4593.} 
\begin{tabular}{lccc} 
\hline \hline
\noalign{\smallskip}
Line    &  FWZI             &   FWHM$_{v=0}$     &   FWHM$_{v=1500}$    \\
        &  [km$\,$s$^{-1}$] &   [km$\,$s$^{-1}$]    & [km$\,$s$^{-1}$]  \\  
\noalign{\smallskip}
\hline 
\noalign{\smallskip}
\ion{O}{i}$\,\,\,\,\,\lambda 8446$   &  6240 $\pm$ 300  &   3830 $\pm$ 100  &   3580 $\pm$ 100\\
\ion{Ca}{ii}$\,\lambda8662$  & 5450 $\pm$ 300    &   4360 $\pm$ 100     &   3680 $\pm$ 100   \\

\hline 
\label{tab:lineWidths} 
\end{tabular} 
\end{table}
%

%
\subsection{Fitting \ion{Ca}{ii}$\,\lambda8662$ with an elliptical accretion disk model}\label{sec:results_fitting}
%

\begin{figure}[h!]
    \centering
    \includegraphics[width=0.46\textwidth,angle=0]{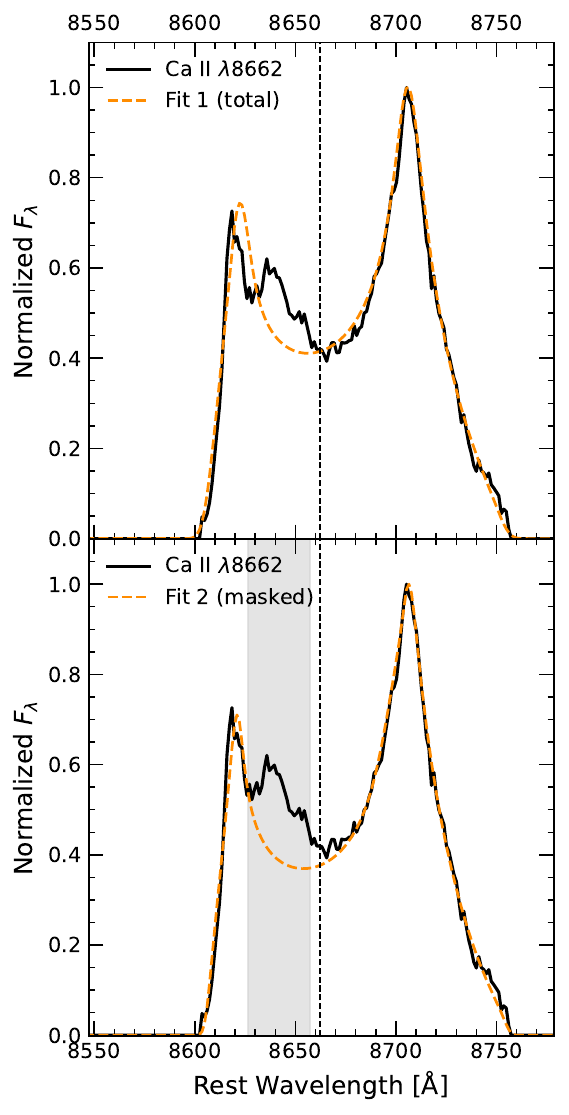}
    \caption{Best-fit elliptical disk-line profiles (orange) for the total (upper panel) and partially masked (lower panel) \ion{Ca}{ii}$\,\lambda8662$ profile. The masked region is shown in gray.}
    \label{fig:OI_CaII_disk_NGC4593}
\end{figure}

To assess whether the observed \ion{Ca}{ii}$\,\lambda8662$ profile is consistent with emission from a non-axisymmetric disk, we fitted  it using the rotating relativistic elliptical disk model of \citet{eracleous95}. We first fitted the full profile (Fit 1) and then a masked version (Fit 2), where part of the blue profile is excluded.  Details of the fitting procedure are provided in Appendix~\ref{sec:app_fitting_CaII}; the best-fit parameters are listed in Table \ref{tab:bestfit_parameters}; and a comparison of Fit 1, Fit 2, and the observed profile is presented in Fig.~\ref{fig:OI_CaII_disk_NGC4593}.

We find the best-fit model (Fit 2) to describe a mildly eccentric ($e \sim 0.22)$, low-inclination ($i \sim 11\degree$) and minimal-turbulence ($\sigma \sim 65$\,\kms{}) disk. The \ion{Ca}{ii} emission region is confined to $\sim 320$-$1100$\,r$_g$. This model accurately reproduces the blue and red peaks, the central dip, and the red foot of the profile.

Unlike previous studies using an elliptical disk model \citep[e.g.,][]{hung20, wevers22, ochmann24}, our fit does not require an additional central component to describe the central profile.  However, the small peaks on the blue side of the profile (see \ref{sec:results_extracting}) are inconsistent with emission from a purely homogeneous elliptical disk and appear to be superimposed on the disk-line profile.

\begin{table*}[ht!]
\caption{Best-fit parameter sets for the \ion{Ca}{ii}\,$\lambda 8662$ profile, for both total and masked profiles, using the disk model of \citet{eracleous95}.}
\centering
\small 
\setlength{\tabcolsep}{4pt} 
\renewcommand{\arraystretch}{1.2} 
\begin{tabular*}{\textwidth}{@{\extracolsep{\fill}}l|ccccrccc@{}}
    \hline\hline
    \noalign{\smallskip}
    Profile & $\xi_1$ [r$_g$] & $\xi_2$ [r$_g$] & $i$ [deg] & $\phi_0$ [deg] & $\sigma$ [\kms{}] & $e$ & $q$ \\
    \noalign{\smallskip}
    \hline
    \noalign{\smallskip}
    Fit 1 (total) & $369^{+14}_{-11}$ & $1305^{+65}_{-29}$ & $11.1^{+0.2}_{-0.2}$ & $65.7^{+5.8}_{-4.5}$ & $114^{+21}_{-18}$ & $0.17^{+0.01}_{-0.01}$ & $-1.19^{+0.12}_{-0.08}$ \\
    Fit 2 (masked) & $320^{+9}_{-10}$ & $1103^{+29}_{-33}$ & $10.8^{+0.1}_{-0.1}$ & $57.1^{+1.6}_{-1.1}$ & $65^{+11}_{-8}$ & $0.22^{+0.01}_{-0.01}$ & $-1.27^{+0.05}_{-0.05}$ \\
    \hline
\end{tabular*}
\label{tab:bestfit_parameters}
\end{table*}

%
\section{Discussion}\label{sec:discussion}
%

%
\subsection{The emission zone and kinematics of the BLR}\label{sec:discussion_kinematics_BLR}
%

The detection of double-peaked NIR \ion{Ca}{ii} triplet profiles in NGC\,4593 represents the first unambiguous case of double-peaked \ion{Ca}{ii} profiles in a nontransient AGN spectrum. Previously, double-peaked \ion{Ca}{ii} profiles had only been observed in a transient AGN spectrum \citep{ochmann24}. The identification of such profiles in a nontransient AGN builds on the systematic search for \ion{Ca}{ii} triplet emission in AGN, initiated by \citet{persson85,persson88b} in the 1980s. Based on the observed \ion{Ca}{ii} triplet ratio of 1:1:1, the emission region of \ion{Ca}{ii} is a high-density zone, in agreement with previous observations and photoionization calculations \citep[e.g.,][]{persson88b, ferland89, joly89}. The disk signature of the \ion{Ca}{ii}$\,\lambda8662$ profile (see \ref{sec:results_fitting}) closely associates this zone with an outer accretion disk, as already suggested by \citet{ferland89} and \citet{dultzin-hacyan99} for other objects. In the case of NGC\,4593, this disk is a non-axisymmetric rotating relativistic disk and the \ion{Ca}{ii} emission region is confined to $\sim 320$-$1100$\,r$_g$, which aligns well with BLR sizes inferred from the analysis of double-peaked Balmer lines \citep[e.g.,][]{strateva03,ward24}.
A disk-like BLR geometry in NGC\,4593, but with a higher inclination angle, was already suggested by \citet{williams18} on the basis of dynamical modeling of the \Hb{} emitting region.

Based on the similar width of \ion{Ca}{ii}$\,\lambda8662$ and \ion{O}{i}$\,\lambda8446$, their identical red-to-blue peak ratio of 4:3, and the matching shape of their red wings, we suggest that \ion{Ca}{ii} and \ion{O}{i} originate from similar and overlapping regions within the BLR, consistent with previous findings \citep[e.g.,][]{rodriguez-ardila02}. However, the shallower blue wing of \ion{O}{i}$\,\lambda8446$ and the presence of additional narrow or intermediate-width components not seen in \ion{Ca}{ii}$\,\lambda8662$ indicate that these regions are not exactly identical. These differences in line profile could also potentially be influenced by optical-depth effects \citep{eracleous09, flohic12}. Given the similar kinematic signatures of \ion{Ca}{ii} and \ion{O}{i} in NGC\,4593, we propose that their emission regions are more closely related to each other than to that of \Hb{}, which exhibits a different red peak position (at $\sim1250$\,\kms{}) and a FWHM$_{v=1250}$ of $\sim 4500$\,\kms{}, which means that it is larger by about 20\% in comparison to \ion{Ca}{ii}$\,\lambda8662$.\footnote{The presence of double-peaked profiles in \Hb{} and \Ha{} is evident in the difference and rms profiles shown by \citet{kollatschny97}.}
A detailed comparison between the Balmer lines and \ion{Ca}{ii} and \ion{O}{i} is, however, beyond the scope of this manuscript.

Although the elliptical disk model accurately reproduces the main features of the \ion{Ca}{ii}$\,\lambda8662$ profile, it cannot account for the smaller peaks in the blue part. These could be caused by the presence of additional BLR components or BLR inhomogeneities, as suggested by \citet{ochmann24} for NGC\,1566. Alternatively, these features may indicate that, despite the elliptical disk model providing an excellent overall fit, the underlying structure is more complex. We argue that \ion{Ca}{ii} could be a valuable tool for refining both line-emitting disk models and BLR models in general (see \ref{sec:discussion_diagnostic tools}).

%
\subsection{The \ion{Ca}{ii} triplet and \ion{O}{i}$\,\lambda8446$ as diagnostic tools}\label{sec:discussion_diagnostic tools}
%
According to the elliptical disk fit (see \ref{sec:results_fitting}), the \ion{Ca}{ii}$\,\lambda8662$ emitting region in NGC\,4593 (i.e., the BLR) is best described by a mildly eccentric, low-inclination disk with an internal turbulence of only $\sigma \sim 65$\,\kms{}. To our best knowledge, this is the lowest value reported for the internal turbulence in the literature to date. Together with the absence of narrow or intermediate-width components, this makes \ion{Ca}{ii} in NGC\,4593 one of the clearest kinematic signatures of a relativistic disk observed in BLR emission lines yet.

To date, double-peaked \ion{Ca}{ii} and \ion{O}{i}$\,\lambda8446$ lines have only been reported for two sources,\footnote{A double-peaked \ion{O}{i}$\,\lambda11287$ profile was reported earlier by \citet{diasdossantos23} for the Seyfert 1 galaxy III Zw 002.} namely NGC\,4593 (this study) and NGC\,1566 \citep{ochmann24}. In both cases the reported internal turbulence is less than $100$\,\kms{}. This is well below the typical internal turbulence of $\sigma \gtrsim 500$\,\kms{} for the Balmer lines widely used in DPE studies. This high internal turbulence smears out the double-peaked profile, in some cases effectively obscuring the kinematic signature of the underlying rotating disk. In contrast, the minimal internal turbulence in \ion{Ca}{ii} and \ion{O}{i} supports the idea that their emission region is associated with a dense emission zone \citep[e.g.,][]{matsuoka07} close to the mid-plane of a BLR disk, where the gas is highly kinematically ordered, and line broadening is primarily due to Keplerian rotation. This makes the \ion{Ca}{ii} triplet and \ion{O}{i}$\,\lambda8446$ lines valuable tools for studying and constraining BLR kinematics in AGN using single-epoch spectra. We propose that studies of \ion{Ca}{ii} could aid in refining line-emitting disk models, resolving model degeneracies, and investigating complex kinematics\footnote{At this point we note the presence of complex \ion{Ca}{ii} triplet and \ion{O}{i}$\,\lambda8446$ emission in IC\,4329A, which has not been previously reported. Complex \ion{Ca}{ii} triplet spectra of other sources can be found in \citet{garcia-rissmann05}.} that may arise in systems such as supermassive black hole binaries (SMBHBs) or multi-SMBHs systems \citep[][]{popovic12}. However, this requires high-quality spectra with a sufficient S/N and resolution, and sufficient spatial resolution to minimize the influence of \ion{Ca}{ii} absorption from the host galaxy. 

The power of the \ion{Ca}{ii} triplet and \ion{O}{i}$\,\lambda8446$ lines as diagnostic tools has already been highlighted by other authors \citep[e.g.,][]{marziani13}. Near-infrared \ion{Ca}{ii} and \ion{O}{i} emission has been widely used for this purpose, with a particular emphasis on photoionization modeling \citep[e.g.,][and previous references]{panda20, panda21, martinez-aldama21}. We suggest that combining  photoionization modeling with detailed line profile studies across a larger sample of sources can provide invaluable information about the BLR structure, particularly regarding the presence or absence of intermediate-width components \citep[e.g.,][]{adhikari18}.

%
\section{Conclusions}\label{sec:conclusions}
%
In this study we presented a detailed analysis of the NIR \ion{Ca}{ii} and \ion{O}{i} emission complex in a MUSE spectrum of NGC\,4593. Using a detailed line profile decomposition, we extracted and analyzed the deblended profiles of \ion{Ca}{ii}$\,\lambda8662$ and \ion{O}{i}$\,\lambda8446$, evaluating them in the context of predictions from non-axisymmetric line-emitting accretion disk models. Our key findings are as follows:

\begin{enumerate}

\item The NIR \ion{Ca}{ii} triplet and \ion{O}{i} emission complex in NGC\,4593 comprises four blended line profiles, each exhibiting an asymmetric double-peaked structure with a red-to-blue peak ratio of 4:3. This marks the first detection of double-peaked \ion{Ca}{ii} triplet and \ion{O}{i}$\,\lambda8446$ in a nontransient AGN spectrum. The \ion{Ca}{ii}$\,\lambda8662$ profile is nicely reproduced by a non-axisymmetric relativistic elliptical disk model, reinforcing its applicability to AGN BLRs.

\item The \ion{Ca}{ii} lines (\ion{Ca}{ii}$\,\lambda8498, \lambda8542, \lambda8662$) exhibit identical double-peaked profiles with a line ratio of 1:1:1, indicating their origin in a dense region of the BLR, likely situated near the underlying accretion flow or the accretion disk. This is consistent with the low internal turbulence of $\sigma \sim$ 65 \kms{} derived from the best-fit line-emitting disk model, and is the lowest value reported for an AGN emission line to date.

\item The \ion{Ca}{ii} triplet profiles lack evidence for a central narrow or intermediate-width component, unlike \ion{O}{i}$\,\lambda8446$, which exhibits a central intermediate-width component. However, the line wings of \ion{O}{i} and \ion{Ca}{ii} are nearly identical in shape, with the \ion{O}{i} peaks shifted slightly to lower velocities compared to \ion{Ca}{ii}. This suggests that \ion{O}{i} and \ion{Ca}{ii} originate from similar, though not identical, regions within the BLR.

\end{enumerate}

The observed \ion{Ca}{ii} and \ion{O}{i} profiles in NGC\,4593 represent some of the clearest double-peaked line profiles reported in AGN to date, offering valuable insights into BLR kinematics. Their kinematic signatures remain largely unaffected by internal turbulence, allowing a more direct interpretation of the underlying dynamics. These profiles provide an important benchmark for refining current line-emitting accretion disk models and resolving model degeneracies. Further investigations of \ion{Ca}{ii} and \ion{O}{i} profiles in other AGN are recommended to better constrain BLR properties and the nature of the underlying accretion flow.

\begin{acknowledgements}
We thank the referee for helpful comments. MWO acknowledges the support of the German Aerospace Center (DLR) within the framework of the ``Verbundforschung Astronomie und Astrophysik'' through grant 50OR2305 with funds from the BMWK. PMW was partially supported by the BMBF through the ErUM program (VLT-BlueMUSE 05A23BAC). Research by DC is partly supported by the Israeli Science Foundation (1650/23). The authors greatly acknowledge support by the DFG grants KO 857/35-1, KO 857/35-2 and CH 71/34-3.

\end{acknowledgements}

\bibliographystyle{aa}
\bibliography{literature}

\appendix

\section{Data reduction}
\label{sec:app_data_reduction}

NGC\,4593  was observed under good conditions with eight science exposures and two offset sky exposures in the middle of each block of four. Observation parameters are given in Table~\ref{tab:spectroscopy_log}.

We downloaded the raw data and all calibrations from the ESO archive and let the MUSE pipeline \citep[v2.10.9,][]{weilbacher20} automatically process everything through the EDPS interface \citep{freudling24}, up to an initial creation of separate datacubes for the individual exposures. The reduction steps included bias correction, lamp flat-fielding, wavelength calibration, geometrical calibration, flat-field corrections using twilight sky exposures, flux calibration including telluric absorption correction, sky subtraction, correction to barycentric velocity, astrometric distortion correction, and cube reconstruction including default cosmic ray rejection. Atmospheric refraction was compensated using the optical module inside the instrument, no additional improvement was necessary.  We optimized the sky subtraction by using a higher spatial sky fraction of 20\% (instead of the default of 10\%) to allow the pipeline to re-fit the telluric emission lines in a sky spectrum of higher S/N in the object exposures after the initial estimate on the offset sky field. We aligned all exposures using Moffat fits to the compact central source in all exposures in the reconstructed $R$-band images, and used the Gaia DR3 position \citep{GaiaDR3} as reference.

Our final cube is an unweighted combination of all eight science exposures, using the \texttt{muse\_exp\_combine} recipe, it has a contiguous coverage of about $8\arcsec\times8\arcsec$ and a wavelength range of 4750.25$-$9349.00\,\AA. The image quality achieved with the AO correction can be estimated from the direct measurement of the FWHM of the central source, we derive 0\farcs19, 0\farcs12, and 0\farcs09, respectively, in the $V$, $R$, and $I$ bands. If the source is not point-like, this would give upper limits on the actual spatial resolution of our data. To estimate the S/N of our spectrum in the $r_\mathrm{a}=0\farcs20$ aperture we divided the spectrum with the propagated error. After applying a correction of the error estimates for missing covariances \citep{gunawardhana20} we see that the spectrum has a $S/N>50$ at all wavelengths and $S/N\gtrsim120$ in the wavelength region of interest around the \ion{Ca}{ii} triplet.

\begin{table}[ht!]
\caption{Log of the MUSE observation from 2019 April 28.}
\centering
\resizebox{0.48\textwidth}{!}{
    \begin{tabular}{llcccl}
        \hline \hline
        \noalign{\smallskip}
        Object & Mod. JD & UT Date & Exp. Time   & Seeing    & Mode    \\
           &         &         &  \mcc{[s]}   & {\small FWHM}      &           \\
        \noalign{\smallskip}
        \hline
        \noalign{\smallskip}
         NGC\,4593  &  58601.16    & 2019-04-28  &   $8\times600$ & 0\farcs66$-$0\farcs95 & NFM \\
        \hline
    \end{tabular}}
\label{tab:spectroscopy_log}
\end{table}

\section{Spectral extraction}
\label{sec:app_spectral_extraction}

\begin{figure}[h!]
    \centering
    \includegraphics[width=0.44\textwidth,angle=0]{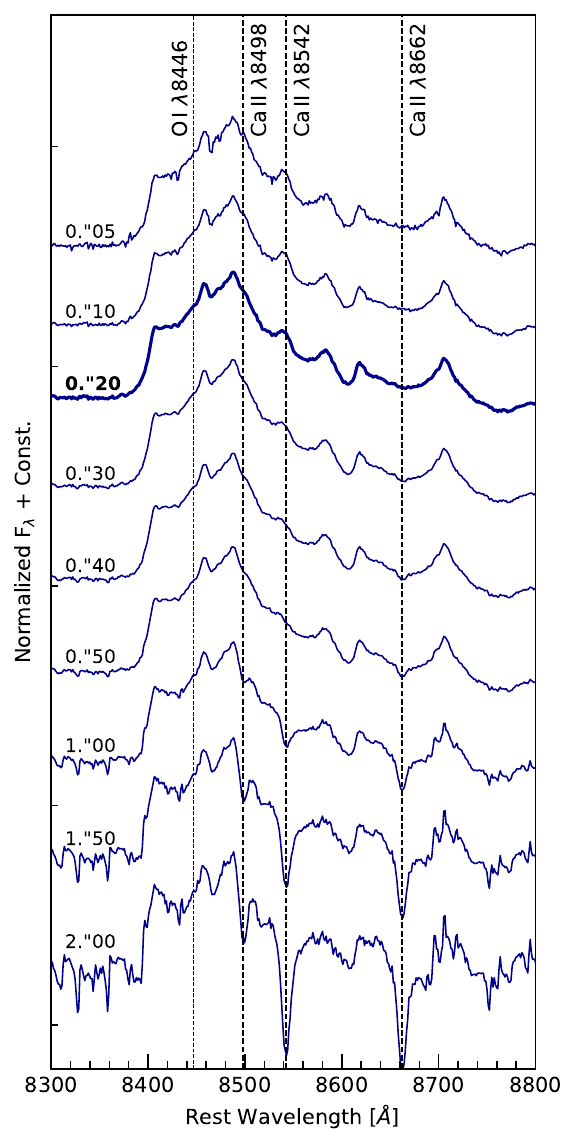}
    \caption{\ion{Ca}{ii} triplet and \ion{O}{i}$\,\lambda8446$ emission complex of the nuclear region for different circular extraction radii $r_{\rm a}$ from the MUSE cube. The spectra are normalized and shifted in flux for clarity. For increasing aperture radii (top to bottom), the influence of stellar \ion{Ca}{ii} triplet absorption becomes stronger. For extraction radii $r_{\rm a}$ larger than 0\farcs50, the host-galaxy contribution starts to dominate, effectively suppressing the nuclear double-peaked \ion{Ca}{ii} triplet and \ion{O}{i}$\,\lambda8446$ profiles. The spectrum with the optimal extraction radius $r_{\rm a}$  of 0\farcs20 is plotted in bold.}
    \label{fig:NGC4593_CaII_absorption}
\end{figure}

For local AGN the \ion{Ca}{ii} triplet and \ion{O}{i}$\,\lambda8446$ spectral range is typically dominated by strong stellar \ion{Ca}{ii} triplet absorption from the host galaxy. This absorption significantly affects the emission line profiles of the \ion{Ca}{ii} triplet lines, making it difficult to assess the genuine shape of the line profiles. For NGC\,4593 , the MUSE observation from 28 April 2019 was obtained in NFM-AO-N, that is, with a central peak $\lesssim$0\farcs09 in the $I$-band. This spatial resolution allows the  extraction of nuclear spectra with very low extraction apertures, thereby reducing the influence of the extended host galaxy. To find the optimal extraction aperture, that is, the aperture for which the S/N of the \ion{Ca}{ii} triplet and \ion{O}{i}$\,\lambda8446$ emission complex is sufficiently high and the influence of the \ion{Ca}{ii} triplet absorption on the emission-line profiles is minimal, we extracted spectra with different aperture radii $r_{\rm a}$ (0\farcs50 up to 2\farcs00) for comparison. The resulting spectra in the rest-frame spectral range from 8300\,\AA{} to 8800\,\AA{} are shown in Fig.~\ref{fig:NGC4593_CaII_absorption}. For larger aperture radii $r_{\rm a}$ the influence of stellar absorption from the host-galaxy becomes stronger and the \ion{Ca}{ii} triplet absorption starts to dominate for $r_{\rm a} \geq$ 0\farcs50, effectively suppressing the nuclear \ion{Ca}{ii} triplet emission lines. For smaller aperture radii $r_{\rm a} \leq$ 0\farcs50 the host-galaxy contribution decreases significantly, effectively vanishing for $r_{\rm a} \leq$ 0\farcs20. We therefore find the spectrum with an aperture radius of $r_{\rm a} =$ 0\farcs20, which translates to $\sim 33$\,pc, to be the optimally extracted spectrum, ensuring minimal host-galaxy contribution and at the same time a sufficient S/N to assess the emission-line profiles. 

\onecolumn
\section{Decomposing the \ion{Ca}{ii} and \ion{O}{i}  blend}
\label{sec:app_decomposing_OI_CaII}

\begin{figure*}[h!]
    \centering
    \includegraphics[width=1.0\textwidth,angle=0]{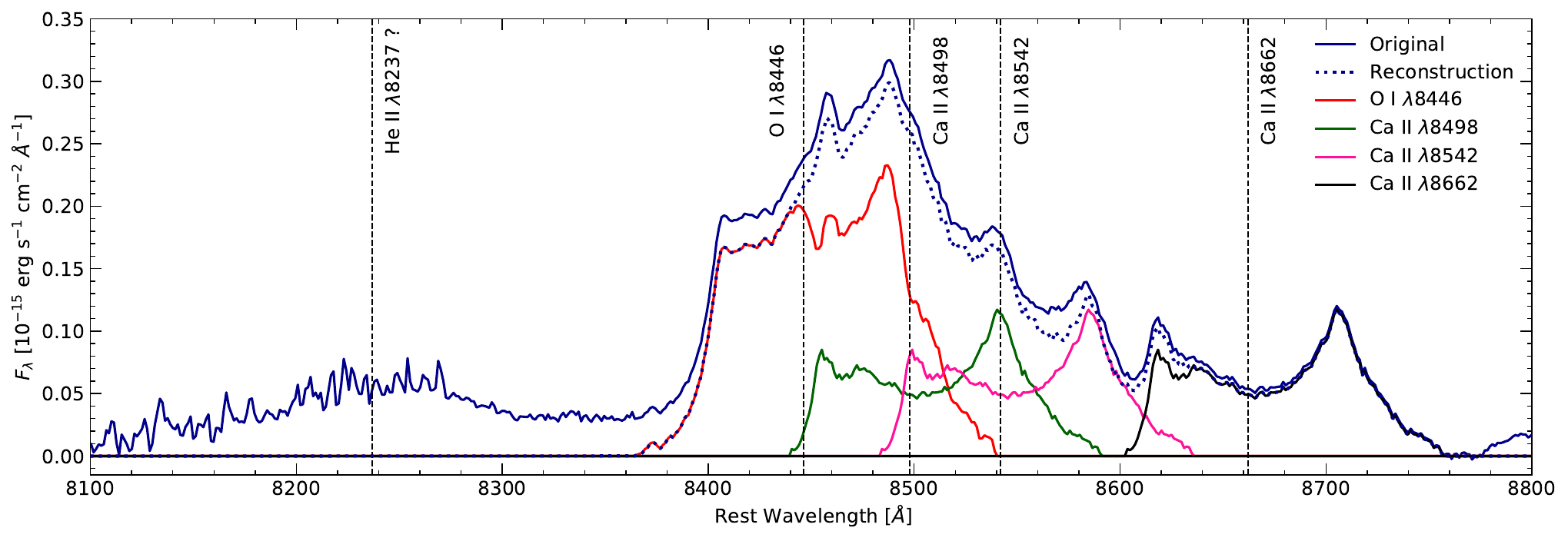}
    \caption{Decomposition of the blended \ion{Ca}{ii} triplet and  \ion{O}{i}$\,\lambda8446$ complex into individual line profiles. The difference between the original spectrum and the reconstructed blend is due to the subtraction of a linear pseudo-continuum beneath each emission line.}
    \label{fig:OI_CaII_complex_NGC4593}
\end{figure*}

The emission complex shown in Fig.~\ref{fig:NGC4593_CaII_absorption} is a blend of the \ion{Ca}{ii} triplet lines \ion{Ca}{ii}$\,\lambda8498,\lambda8542,\lambda8662$, and \ion{O}{i}$\,\lambda8446$. The left wing of \ion{O}{i}$\,\lambda8446$ and the central profile as well as the right wing of \ion{Ca}{ii}$\,\lambda8662$ are the only segments not affected by line blending. The overall shape of \ion{Ca}{ii}$\,\lambda8662$ indicates the presence of a double-peaked profile. To obtain un-blended profiles of the \ion{Ca}{ii} triplet and \ion{O}{i}$\,\lambda8446$ lines, we performed a line-profile decomposition based on two assumptions: First, we assumed that the individual \ion{Ca}{ii} lines have identical line profiles in velocity space, and second, that the ratio between the \ion{Ca}{ii} lines is 1:1:1. Both assumptions are well supported by results from past studies on \ion{Ca}{ii} triplet emission in AGN \citep[e.g.,][]{persson88b, ferland89}. Based on these assumptions, we shifted the central and red part of the \ion{Ca}{ii}$\,\lambda8662$ profile such that its shifted central wavelength matches with the rest-frame central wavelength of \ion{Ca}{ii}$\,\lambda8542$. In this way, we mimicked the central and red part of the \ion{Ca}{ii}$\,\lambda8542$ profile as a shifted \ion{Ca}{ii}$\,\lambda8662$ profile. By subtracting this synthetic \ion{Ca}{ii}$\,\lambda8542$ profile from the \ion{O}{i}$\,\lambda8446$ and \ion{Ca}{ii} triplet line blend (in velocity space), we obtained the clean \ion{Ca}{ii}$\,\lambda8662$ profile used for further an\-alys\-is after subtraction of a linear pseudo-continuum. This clean \ion{Ca}{ii}$\,\lambda8662$ profile was then used as a template for the \ion{Ca}{ii}$\,\lambda8498,\lambda8542$ lines to obtain a clean \ion{O}{i}$\,\lambda8446$ profile by subtracting all \ion{Ca}{ii} triplet lines from the \ion{O}{i}$\,\lambda8446$ and \ion{Ca}{ii} triplet blend. The final decomposition is shown in Fig.~\ref{fig:OI_CaII_complex_NGC4593}. Based on the assumption of identical \ion{Ca}{ii} profiles with a ratio of 1:1:1, we achieved a reconstruction of the \ion{Ca}{ii} + \ion{O}{i} blend that accounts for all features of the blend. The difference between original spectrum and reconstructed blend is due to the subtraction of a linear pseudo-continuum beneath each emission line.

\section{Fitting the \ion{Ca}{ii}$\,\lambda8662$ profile}
\label{sec:app_fitting_CaII}

We modeled the \ion{Ca}{ii}$\,\lambda 8662$ line profile using the elliptical accretion disk model of \citet{eracleous95}. We prefer this model over others, such as disks with spiral structures or warped disks, because it employs the fewest free parameters. This allows us to assess the profile for general consistency with non-axisymmetric models while minimizing complexity. The model includes seven free parameters: the inner and outer pericenter distances ($\xi_1$ and $\xi_2$), the inclination angle ($i$), the orientation of the major axis ($\phi_0$), the velocity broadening parameter ($\sigma$), the disk eccentricity ($e$), and the emissivity power-law index ($q$). To determine the best-fit parameter set for the observed \ion{Ca}{ii}$\,\lambda8662$ profile, we derived posterior probability distributions using the nested sampling Monte Carlo algorithm MLFriends \citep{buchner16, buchner19}, as implemented in the UltraNest\footnote{\url{https://johannesbuchner.github.io/UltraNest/}} package \citep{buchner21}. The an\-alys\-is employed 1000 live points, a slice sampler with 100 steps, and a uniform prior distribution of $50\,$r$_g \leq \xi_1 \leq 20\,000\,$r$_g$, $50\,$r$_g \leq \xi_2 \leq 20\,000\,$r$_g$ (with r$_g$ being the gravitational radius defined as r$_g = GM/c^2$), $0^{\circ} \leq i \leq 60^{\circ}$, $0^{\circ} \leq \phi_0 < 360^{\circ}$, $10\,$\kms{} $ \leq \sigma \leq 3000\,$\kms{}, $0 \leq e < 1$, and $ 0 \leq |q| \leq 6$. The prior was chosen such that it reflects a reasonable parameter space based on results of previous DPE studies \citep{eracleous03, strateva03, storchi-bergmann17, ward24}.

\begin{figure*}[h!]
    \centering
    \includegraphics[width=1.0\textwidth,angle=0]{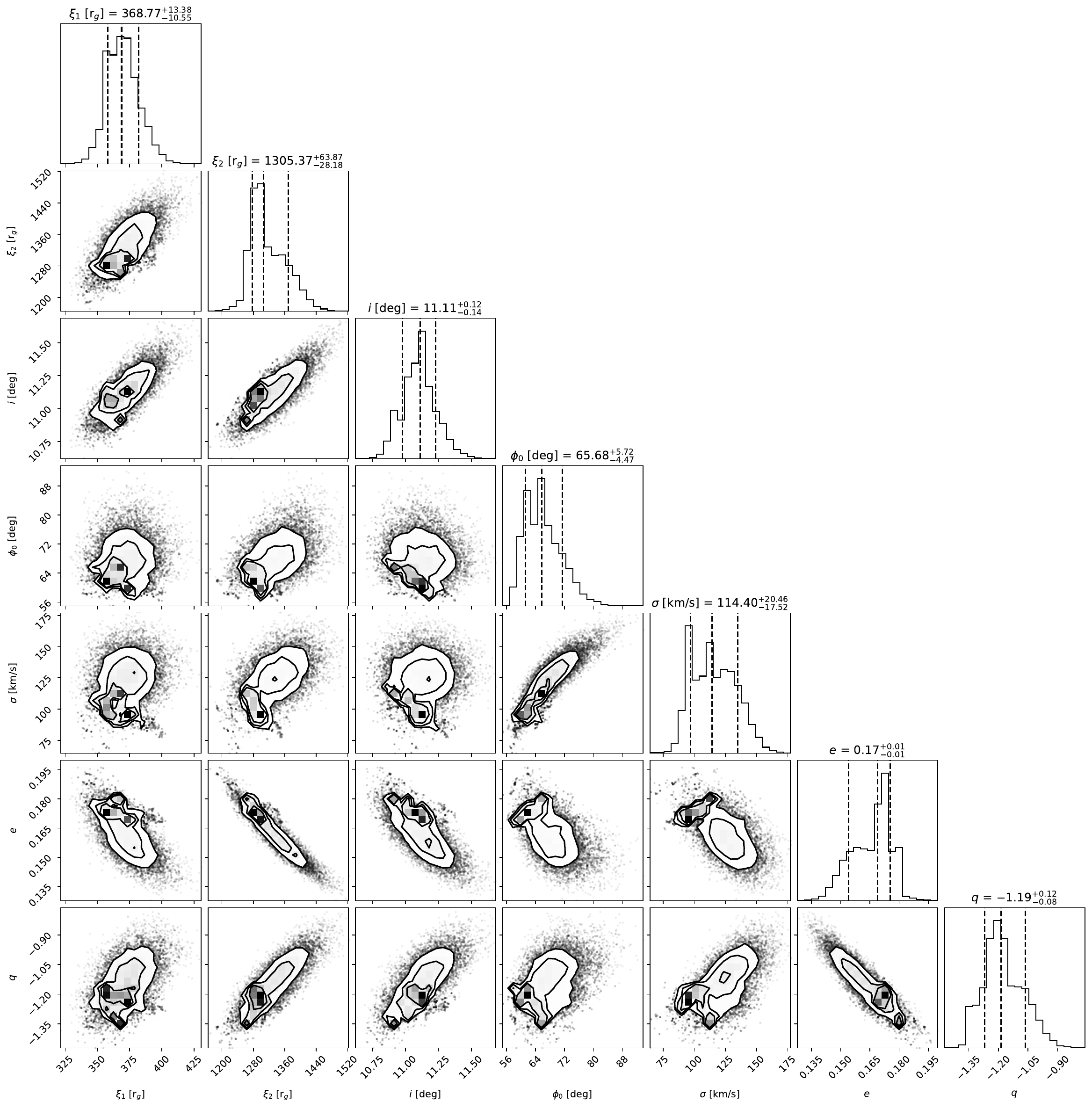}
    \caption{Posterior distributions of the elliptical disk model model parameters for the total profile (Fit 1). The contours correspond to the 16th, 50th, and 84th percentiles. The vertical dashed lines correspond to the best-fit values quoted in Table \ref{tab:bestfit_parameters}.}
    \label{fig:cornerplot_total_NGC4593}
\end{figure*}

\begin{figure*}[h!]
    \centering
    \includegraphics[width=1.0\textwidth,angle=0]{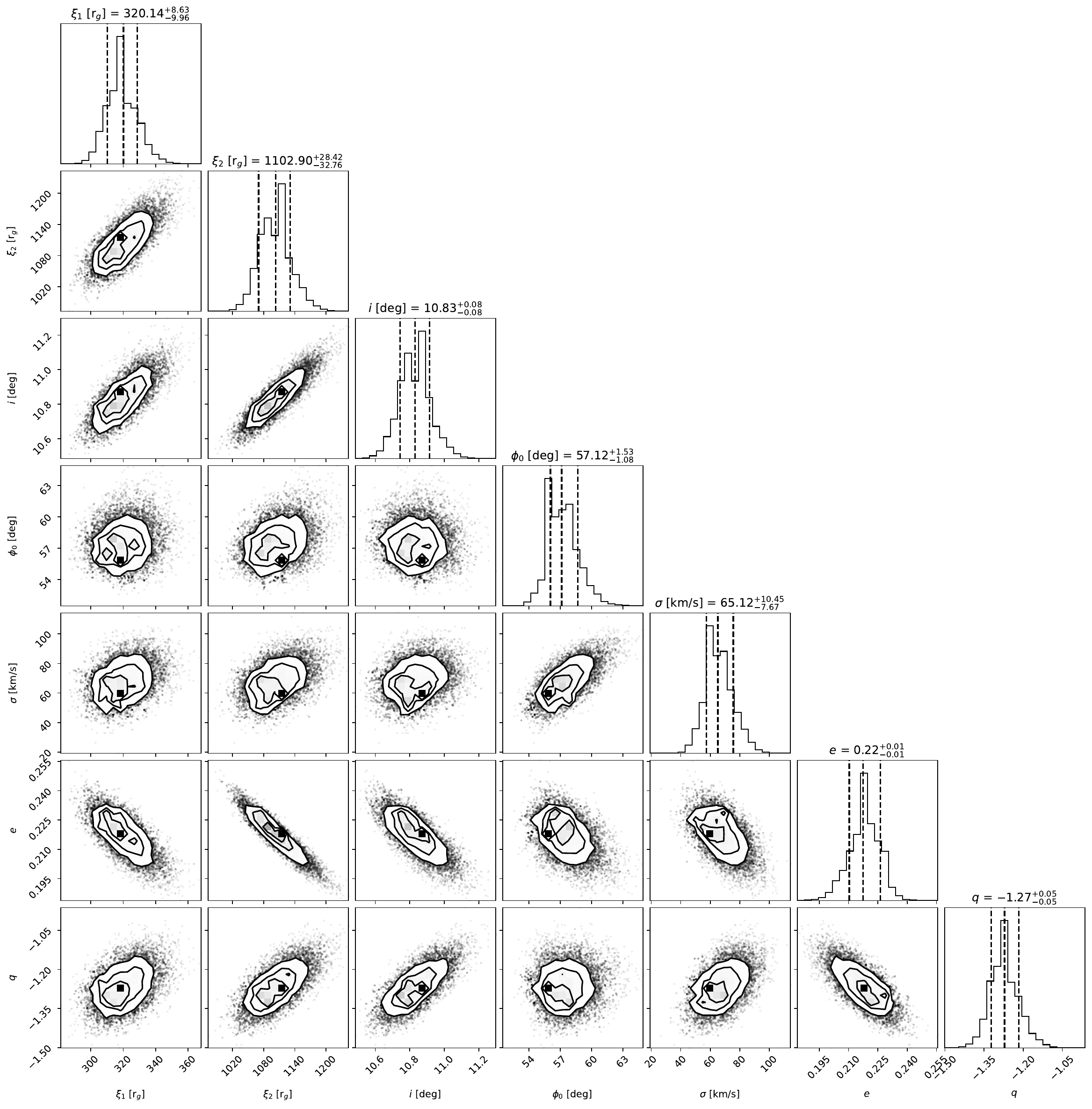}
    \caption{Posterior distributions of the elliptical disk model model parameters for the masked profile (Fit 2). The contours correspond to the 16th, 50th, and 84th percentiles. The vertical dashed lines correspond to the best-fit values quoted in Table \ref{tab:bestfit_parameters}.}
    \label{fig:cornerplot_masked_NGC4593}
\end{figure*}

We first fitted the total profile (Fit 1; see Fig.~\ref{fig:OI_CaII_disk_NGC4593}) and find the best-fit solution to be a mildly eccentric, low-inclination, minimal-turbulence disk. The best-fit parameters are listed in Table \ref{tab:bestfit_parameters}. While this fit reproduces the red part of the profile well, it fails to fully account for the blue part, particularly the blue peak. To address this, we masked the region between $-1260\,$\kms{} and $-140\,$\kms{}, assuming the additional sub-peaks are inconsistent with a homogeneous line-emitting disk model, and fitted only the rest of the profile (Fit 2). The resulting best-fit parameters are listed in Table \ref{tab:bestfit_parameters}. This fit better captures both peaks and successfully reproduces the observed red-to-blue peak ratio of 4:3. The resulting posterior distributions for Fit 1 (total profile) and Fit 2 (masked profile) are shown in \ref{fig:cornerplot_total_NGC4593} and \ref{fig:cornerplot_masked_NGC4593}, respectively. We note that the slight mismatch between the models and the observed blue peak of the \ion{Ca}{ii} profile may be partially due to contamination by [\ion{Fe}{ii}]$\,\lambda8617$, which is not explicitly included in the fitting model.


\end{document}